\documentclass[useAMS,usenatbib]{mn2e}

\usepackage{natbib}
\usepackage{graphicx}
\usepackage{amssymb}

\title[]{How the central black hole may shape its host galaxy through AGN feedback}
\author[W. Ishibashi $\&$ A. C. Fabian]
{W. Ishibashi$^{1,}$$^{2}$\thanks{E-mail:
wako.ishibashi@phys.ethz.ch} and A. C. Fabian$^{1}$ 
\footnotemark[0]\\
$^{1}$Institute of Astronomy, Madingley Road, Cambridge CB3 0HA 
\footnotemark[0]\\
$^{2}$Institute for Astronomy, Department of Physics, ETH Zurich, Wolfgang-Pauli-Strasse 27, CH-8093 Zurich, Switzerland \\ }

\begin{document}

\pdfminorversion=4

\date{Accepted 2014 April 2.  Received 2014 March 12; in original form 2014 January 30}

\pagerange{\pageref{firstpage}--\pageref{lastpage}} \pubyear{2012}

\maketitle

\label{firstpage}

\begin{abstract}
Active galactic nucleus (AGN) feedback provides the link between the central black hole and its host galaxy. 
We assume AGN feedback driven by radiation pressure on dust, which sweeps up the ambient dusty gas into an outflowing shell, and consider feedback-triggered star formation in the outflow. 
An upper limit to the characteristic size of galaxies may be defined by the critical radius beyond which radiation pressure on dust is no longer able to drive the shell. The corresponding enclosed mass may be compared with the host galaxy bulge mass. We show that the resulting relation between characteristic radius and mass, of the form $R \propto \sqrt M$, corresponds to the observed mass-radius relation of early-type galaxies. 
We suggest that such simple physical scalings may account for a number of observed galaxy scaling relations. 
In this picture, both the size and structural evolution of galaxies can be interpreted as a consequence of AGN feedback-driven star formation, mainly associated with the spheroidal component.
The accreting black hole is responsible for triggering star formation in the host galaxy, while ultimately clearing the dusty gas out of the host, thus also contributing to the chemical evolution of galaxies. 
We discuss the importance of radiation pressure on dust in determining the galaxies large-scale properties, and consider the possibility of the central black hole directly shaping its host galaxy through AGN feedback. 
\end{abstract}


%%%%%%%%%%%%%%%%%%%%%%%%%%%%%%%

\section{Introduction}

Observations over the past decades have revealed important connections between the central supermassive black hole and its host galaxy \citep[see the recent review by][and references therein]{Kormendy_Ho_2013}. 
The observed correlations suggest a close coupling, supporting the so-called black hole-galaxy `co-evolution' scenario. 
However, the difference in physical scales between the central black hole and the host galaxy is huge; and at first sight, one would not expect any causal connection between the black hole and the galaxy-scale star formation. 
On the other hand, the energy released by the accreting black hole can easily exceed the binding energy of the galaxy bulge. 
The key question is how the tiny black hole at the centre can influence, and possibly even determine, the fate of an entire galaxy. 

It is now widely agreed that the required link is provided by some form of `feedback' from the central active galactic nucleus (AGN), and different AGN feedback models have been proposed in the literature \citep[][and references therein]{Silk_Rees_1998, Fabian_1999, King_2003, Murray_et_2005, King_2010, Fabian_2012}. 
Feedback can operate in both radiative and kinetic modes, via jets, winds, and radiation pressure.
In the context of galaxy evolution, AGN feedback is most generally invoked to suppress star formation in the host galaxy, either by removing or heating the ambient gas. 
This is the well established negative feedback scenario in which most of the past studies have been discussed. 

However, feedback from the central black hole may also play other roles in galaxy evolution. Different forms of positive feedback have been considered in the past. 
For instance, triggering of star formation due to radio jet activity has been proposed to explain the alignment of radio and optical structures observed in high redshift radio galaxies \citep[][]{Rees_1989, Begelman_Cioffi_1989}.  
Based on these early models, \citet{Chokshi_1997} suggested that jet-triggered star formation may be responsible for the formation of elliptical galaxies, with the most powerful jets associated with giant ellipticals and lower power jets linked to smaller spheroids and central bulges. AGN jet-induced star formation, leading to enhanced star formation rates, has also been considered as a source powering luminous starbursts \citep{Silk_2005}.
More recently, numerical simulations of radio jet-triggered star formation in gas-rich disc galaxies have been performed \citep{Gaibler_et_2012}. 
We also note that `mixed' scenarios, involving both positive and negative feedback, have been recently debated in the literature \citep{Silk_2013, Zubovas_et_2013, Zinn_et_2013}. 

We have previously studied the possibility of AGN feedback triggering star formation in the host galaxy \citep{Paper_1}, 
and suggested that this particular form of galaxy growth may be linked to the observed size evolution of massive galaxies over cosmic time \citep{Paper_2}. 
Indeed, numerous observational studies show that massive, quiescent galaxies at high redshift ($z \sim 2$) are much smaller than galaxies of comparable mass in the local Universe, implying a significant size evolution over the past $\sim$10 Gyr \citep[e.g.][]{Bezanson_et_2009, vanDokkum_et_2010, Patel_et_2013}. The observed increase in size seems to follow a characteristic `inside-out' growth pattern, whereby most of the growth takes place at large radii, leading to the gradual build-up of an outer envelope. 
Observations also indicate that, coupled with the increase in radius, significant structural and morphological changes occur over the same time span, as seen by variations in the Sersic index \citep{vanDokkum_et_2010, Patel_et_2013, Buitrago_et_2013}.  
This suggests a parallel evolution in the size and structure of the growing galaxies, which need to be simultaneously accounted for.  

Here we explore whether the galaxies overall evolution can be interpreted within a single physical framework in which the central black hole plays a major role. In particular, we examine how the accreting black hole may determine the characteristic properties of its host galaxy through AGN feedback driven by radiation pressure on dust.

%%%%%%%%%%%%%%%%%%%%%%%%%%%%%%%%%

\section{Feedback-driven star formation}

We briefly summarise here the main features of the AGN feedback-driven star formation model, discussed in more detail in \citet{Paper_1}. 
We consider the direct effects of radiation pressure on dusty gas as it sweeps up the ambient material into an outflowing shell. 
The general form of the equation of motion of the shell is given by: 
\begin{equation}
\frac{d}{dt} [M_\mathrm{g}(r) \dot{r}] = \frac{L}{c} - \frac{G M_\mathrm{g}(r) M_\mathrm{DM}(r)}{r^2} 
\label{Eq_motion}
\end{equation} 
where $L$ is the luminosity of the central source, $M_g(r)$ the enclosed gas mass, and $M_{DM}(r)$ is the dark matter mass. 
The same equation of motion for the case of a shock pattern propagating in the host galaxy has been analysed by \citet{King_2010}. 
Assuming an isothermal potential, we obtain that the shell velocity always tends towards a constant asymptotic value at large radii:  
\begin{equation}
v_\mathrm{\infty} = \sqrt{ \left( \frac{G L}{2 f_\mathrm{g} c \sigma^2} - 2 \sigma^2 \right) } 
\end{equation} 
where $\sigma$ is the velocity dispersion and $f_g$ the gas mass fraction. 

The interstellar medium of the host galaxy is swept up by the passage of the expanding shell. 
The resulting squeezing and compression of the ambient medium induce local density enhancements, which in turn can trigger star formation within the outflowing shell.  
The star formation rate in the outflow is parametrised as 
\begin{equation}
\dot M_{\star} \sim \epsilon_{\star} \,  \frac{M_\mathrm{g}(r)}{t_\mathrm{flow}(r)} \sim \epsilon_{\star} \,  \frac{2 f_\mathrm{g} \sigma^2}{G} v(r)
\end{equation} 
where $\epsilon_{\star}$ is the star formation efficiency and $t_\mathrm{flow}(r)$ is the local flow time. 
We see that higher star formation rates are expected for more massive black holes, provided that the star formation efficiency is a constant independent of the black hole parameters. 
Empirical star formation laws suggest that this efficiency is typically of the order of a few percent in star-forming galaxies, with possibly higher values in ultraluminous starbursts at high redshifts \citep{Genzel_et_2010}. 
We note that the star formation efficiency is not constrained by any theory and remains a free parameter in the model. 

The newly formed stars are initially expected to follow radial orbits; but due to the effects of violent relaxation, the stellar orbits are likely to be soon rearranged, settling down on a timescale comparable to the crossing time. 
The formation of new stars at increasingly larger radii in the outflow gradually builds-up the outer regions of the host galaxy, leading to a substantial increase in the galaxy's size. 
We have argued that several such episodes of star formation, due to AGN feedback activity, may account for the observed size evolution of massive galaxies \citep{Paper_2}. 

%%%%%%%%%%%%%%%%%%%%%%%%%%%%%%%%%%%

\section{Radiation pressure on dust and galaxy scaling relations}

Galactic-scale AGN feedback can be driven by different physical mechanisms, such as high-velocity winds \citep[e.g.][]{King_2010} and radiation pressure on dust (see \citet{Paper_1} for a discussion comparing the two driving mechanisms). 
In contrast to fast winds, which generate shockwaves propagating into the interstellar medium, radiation pressure acts on the bulk of the mass and we do not expect strong shocks to develop. 
We assume that radiative feedback mainly operates on dense, dusty gas, which is strongly radiating and rapidly cools down. 
Indeed, radiative cooling of the dense gas is very efficient and ensures that the outflowing matter remains cold (except possibly for the development of an outer shock in the swept-up gas, which subsequently cools down). In the following, we focus on the direct effects of radiation pressure on dusty gas.

\subsection{The critical radius}

We recall that the key element here is radiation pressure on dust. 
In the inner regions, radiation from the central source is efficiently absorbed by dust grains and the ambient dusty gas is swept up into a shell. The optical depth of the expanding shell is given by
\begin{equation}
\tau \sim \frac{\kappa M_g(r)}{4 \pi r^2} \sim \frac{\kappa f_g \sigma^2}{2 \pi G r} \end{equation}
where $\kappa$ is the medium opacity. 
As the shell expands, the optical depth gradually falls off, and photons start to freely escape. 
We can define a critical radius, $R_c$, where the outflow becomes optically thin to the central radiation, i.e. where the optical depth drops to unity ($\tau(R_c) \sim 1$):
\begin{equation}
R_c \sim \frac{\kappa f_g \sigma^2}{2 \pi G} 
\label{Eq_Rc}
\end{equation}
Beyond this critical radius, radiation pressure-driving becomes inefficient and is no longer able to push the shell: the sweeping up of dusty gas is effectively halted at this point. 
The precise location of the critical radius depends primarily on the coupling process assumed, i.e. electron scattering or dust absorption.  
We recall that the effective Eddington limit for dust differs from the standard Eddington limit by a factor $\sigma_d/\sigma_T$, where $\sigma_d$ is the dust absorption cross section and $\sigma_T$ is the Thomson cross section, with the ratio of the order of $\sigma_d/\sigma_T \sim 10^3$ \citep[e.g.][]{Fabian_2012}. 
For typical dust opacities ($\kappa_d \sim 500 \, \mathrm{cm^2 g^{-1}}$), the corresponding critical radius is of the order of a few tens of kiloparsecs: $R_c \sim 25 \, \kappa_{d, 500} f_{g, 0.16} \sigma_{200}^2$ kpc. 
The critical radius may thus define a physical boundary, which sets an upper limit to the characteristic size of galaxies. 

The mass enclosed within the critical radius, assuming an isothermal distribution, is given by
\begin{equation}
M_c(R_c) \sim \frac{2 \sigma^2}{G} R_c \sim \frac{f_g \kappa_d \sigma^4}{\pi G^2}
\end{equation}
where $\kappa_d = \sigma_d/m_p$. 
This corresponds to the galaxy mass
\begin{equation}
M_{gal} \sim \frac{f_g \sigma_d \sigma^4}{\pi G^2 m_p}
\label{Eq_galmass}
\end{equation}
obtained by considering the effective Eddington limit and assuming that successive cycles of radiation pressure on dust are responsible for the observed $M_{BH}-\sigma$ relation \citep[][]{Fabian_2012}.
In fact, considering a sequence of ejection and re-accretion episodes in a cycling scenario, the ratio of bulge mass to black hole mass tends towards $M_{b}/M_{BH} \sim \sigma_d/\sigma_T \sim 10^3$, which is quite close to the observed ratio \citep{Marconi_Hunt_2003}. This also corresponds to the ratio of critical radii, $R_c/R_{c,e} \sim \sigma_d/\sigma_T$, where $R_{c,e}$ is the critical radius for electron scattering. 
The electron scattering opacity dominates in the vicinity of the central AGN, where dust grains are heated beyond their sublimation temperature and effectively destroyed.
But sublimation of dust is only relevant in the innermost regions (within the sublimation radius on $\sim$pc-scales, cf \citet[][]{Murray_et_2005}), and the dust opacity becomes dominant on larger scales, where galaxy-wide outflows are observed. 
Therefore, while the local physics in the immediate vicinity of the central black hole may be determined by electron scattering processes, the global properties on galactic scales may be governed by dust interaction processes. 
In the particular form of radiative feedback envisaged here, galaxies cannot grow indefinitely and their physical extent is ultimately limited by the dust content. 
The galaxies characteristic properties, such as size and mass, could then be essentially set by the effects of radiation pressure on dust. 

In the above discussion, we have considered a simple homogeneous expansion, while in more realistic cases, the dusty gas may be condensed into dense clumps within the outflowing shell. This may lead to a lowering of the global dust covering fraction and reduction of the radiative momentum transfer. On the other hand, new dust can be produced when massive stars (formed within the feedback-driven outflow) explode as supernovae. 
In fact, recent Herschel observations of the Crab nebula suggest that dust production in core-collapse supernovae can be quite efficient \citep{Gomez_et_2012}. 
Fresh dust can then be re-injected and spread into the surrounding medium, enhancing the overall feedback process. 

%---------------------------------------------------------

\subsection{The mass-radius relation and connections to the Fundamental Plane}

The predicted scaling of the critical radius with velocity dispersion (Eq. \ref{Eq_Rc}) coupled with the $M_{BH} - \sigma$ relation implies a relation of the form: $R_c \propto \sigma^2 \propto \sqrt{M_{BH}}$. For a constant mass fraction ($M_{BH}/M_b \sim \sigma_T/\sigma_d$), this yields $R_c \propto \sqrt{M_b}$.
Thus, there seems to be a well defined relation between characteristic radius and mass of the form $R \propto M^{1/2}$. 
Interestingly, this corresponds to the observed mass-radius relation of early-type galaxies. 
For instance, \citet{Shen_et_2003} obtain $R \propto M^{0.55}$ for local early-type galaxies based on a large sample from the Sloan Digital Sky Survey (SDSS).  
Massive galaxies at $z \sim 1.4$ are also found to follow a size-mass relation with the same slope, but offset in normalisation compared to the local relation \citep{McLure_et_2013}. 
The offset requires a growth factor of a few, consistent with the size evolution inferred from other studies.  
Although a significant evolution in size is observed over cosmic time, there seems to be not much evolution in the slope of the mass-radius relation with redshift \citep{Newman_et_2012}.  
This is confirmed by a recent study based on a sample of spectroscopically confirmed massive galaxies at $z \sim 2$, which is observed to follow a mass-radius relation with a slope of $\sim 0.51 \pm 0.32$, consistent with the local relation \citep{Krogager_et_2013}.

Assuming a constant mass-to-light ratio, the scaling of galaxy mass with velocity dispersion (Eq. $\ref{Eq_galmass}$) can also be considered as analogous to the Faber-Jackson relation, which is followed by the spheroidal component of galaxies \citep{Fabian_2012}. 
Furthermore, combining the scaling of the critical radius ($R_c \propto \sigma^2$) with the $M_{BH} -\sigma$ relation ($M_{BH} \propto \sigma^4$) may even account for the observed black hole Fundamental Plane, of the form $M_{BH} \propto \sigma^3 R^{1/2}$ \citep{Hopkins_et_2007}.

In the innermost regions close to the central black hole, Compton heating may prevent gas cooling and locally suppress star formation. 
The heating process is important only within a certain radius, which depends on the central luminosity and hence black hole mass, with the Compton heated zone extending to a larger radius for a more massive object.
As a consequence, for a larger galaxy having a larger central black hole, the star formation suppression zone will be more extended. 
On the other hand, a large black hole mass also implies higher star formation rates at outer radii, resulting in a larger effective radius. 
The overall trend may be compared with the Kormendy relation, which indicates that larger elliptical galaxies have lower surface brightness. 
Comparing the mean surface brightness within the effective radius of massive elliptical galaxies at $z \sim 1.5$ with the corresponding values of counterparts at low redshifts, \citet{Longhetti_et_2007} concluded that an increase in the effective radius is required in order to match the local Kormendy relation. This has also been pointed out as a further evidence for the size evolution of massive galaxies.

%%%%%%%%%%%%%%%%%%%%%%%%%%%%%%%%%

\section{Structural and morphological evolution}

We have previously discussed the aspect of the size growth of galaxies due to AGN feedback-driven star formation \citep{Paper_2}. 
As the increase in radius is not a uniform process, we also expect associated structural changes, which might affect the morphology of the growing galaxies. 
In fact, the formation of new stars at outer radii in the feedback-driven outflow results in a significant change in the surface density profiles, which observationally can be quantified by variations in the Sersic index. 
Indeed, the development of extended outer envelopes surrounding the inner core, associated with the inside-out growth, is thought to be responsible for the increase of the Sersic index observed in massive galaxies \citep{vanDokkum_et_2010, Patel_et_2013, Buitrago_et_2013}. 
The large Sersic values observed in local ellipticals are usually attributed to their bright outer envelopes and is indicative of a spheroidal geometry \citep{Buitrago_et_2013}, also confirmed by the observed axis ratio \citep{Patel_et_2013}. 
Feedback-triggered star formation occurring along the minor axis of a galaxy may also contribute to the growth of the spheroidal component, as recently observed in a high-redshift radio galaxy \citep{Hatch_et_2013}. 
In general, since AGN-driven outflows are wide-angle phenomena, and the associated star formation is presumably not confined in a single plane, we may speculate that feedback-driven star formation is mainly linked to the spheroidal component of galaxies.

The spatial extent of the outer envelope formed by feedback-triggered star formation depends on the radial distance reached by the AGN outflow. 
Within a given AGN activity time scale, a typical radius can be of the order of $R_{\infty} \sim v_{\infty} \Delta t_{AGN}$, which ultimately depends on the central luminosity and hence black hole mass. 
A massive black hole will be able to drive the shell out to large radii and thus affect the host on scales of the entire galaxy, while a smaller black hole will only push the shell to smaller radii and only affect the inner regions of the host. 
Assuming that the feedback timescale is comparable to the Salpeter time, the characteristic radii can range from $\sim$kpc- to $\sim$tens of kpc- scales, depending on the central black hole parameters. 
These size scales roughly correspond to the typical size of the central bulge in spiral galaxies and of the whole bulge in elliptical galaxies, respectively. 
We note that, in many cases, the sizes of high-redshift compact galaxies are found to be smaller than the sizes of the bulges of present-day spirals. 

The most massive black holes would then be connected with large elliptical galaxies, while smaller black holes may be linked to the central bulges of disc galaxies. 
A similar conclusion has been reached by \citet{Zubovas_King_2012b} on different grounds; namely that the outflow removal time in an elliptical galaxy is longer than that for clearing the bulge of a spiral galaxy, implying that the final black hole mass in red and dead ellipticals is larger than that in spiral galaxies. 
Observations show that the most massive black holes are indeed associated with giant ellipticals, while lower mass black holes reside in late-type disc galaxies \citep[see][for a recent compilation]{McConnell_Ma_2013}. 
It is interesting to note that the observational correlations between the central black hole and its host galaxy only concern the spheroidal component, i.e. the entire galaxy in the case of ellipticals, and the central bulge in the case of spirals; whereas there is little or no connection with the disc component \citep[][and references therein]{Kormendy_Ho_2013}. 
This suggests that bulges and discs have distinct physical origins, with the central black hole only related to the spheroidal component of galaxies. 

Following the episode of star formation triggering, we expect radiation pressure to sweep away the remaining gas and dust.  
Hence AGN can not stay excessively dusty, since much of the dusty gas is eventually removed from the host.
The galaxy depleted of its gas and dust content may thus appear as a red and dead elliptical remnant. 
However, gas evacuated from the core of the galaxy might still be present in the galaxy outskirts, retained in the outer dark halo.
In fact, recent observations indicate the presence of substantial amounts of cool gas in the outer regions of early-type galaxies and quasar host haloes \citep{Thom_et_2012, Prochaska_et_2013}.
 
In the case of a less powerful central source, an analogous process can take place on reduced scales. 
Gas may be removed from the nuclear region, leaving a depleted inner bulge, reminiscent of the red central bulge of spiral galaxies. 
The displaced gas may not be expelled from the main body of the galaxy, but instead remain piled up at modest radii, as seen in some numerical simulations \citep{Debuhr_et_2011}. 
The dusty gas swept out and deposited in the galaxy outskirts is potentially available to fuel the next cycle of re-accretion and feedback episodes, leading to some metal enrichment. 
Observationally, the dust content seems to be essentially determined by the star formation activity, with no clear evolution of the dust fraction with redshift for a given star formation rate and stellar mass \citep{Santini_et_2014}. 
In our case, there is a complex coupling between star formation activity and dust evolution due to the common underlying feedback mechanism, with dusty gas being driven out in the outflow. Radiation pressure on dust could therefore have an influence on the chemical evolution of galaxies, and contribute to the metal enrichment of the intergalactic medium.

%%%%%%%%%%%%%%%%%%%%%%%%%%%%

\section{Discussion}

Recent observations have reported detections of galactic-scale molecular outflows in a number of active sources \citep[e.g.][]{Sturm_et_2011, Cicone_et_2014}. The measured outflow parameters (velocities, kinetic powers, momentum rates) seem to be consistent with model predictions based on energy-driven flows \citep{Zubovas_King_2012a}, thus favouring energy-driving over momentum-driving. 
However, the observed large momentum boosts ($\dot{M} v \sim 10 L/c$) do not necessarily rule out radiation pressure driving, since matter can be optically thick to reprocessed infrared radiation, leading to momentum fluxes reaching several times the single scattering limit \citep{Debuhr_et_2011, Roth_et_2012}. 
On the other hand, it has been argued that the coupling between matter and radiation can be inefficient in a range of contexts, due for instance to instabilities developing in the ambient medium \citep{Faucher-Giguere_Quataert_2012, Krumholz_Thompson_2013}. 
A clearer picture depends on the details of multi-dimensional effects, and demands a better understanding of radiative physics. 
In any case, we consider radiation pressure on dust as a viable mechanism for driving large-scale outflows, although the most extreme examples may require other acceleration processes. 
One should also keep in mind that the observational samples are often biased towards powerful sources with previously known outflow detections \citep{Cicone_et_2014}. 

Various physical mechanisms aimed at explaining the observed size and structural evolution of galaxies have been discussed in the literature. 
In general, both the size and structural changes are interpreted in terms of external processes, such as a sequence of merger events. At present, some form of `two-phase' galaxy assembly, with an initial in-situ star formation epoch followed by a later phase of satellite accretion, seems to emerge as the favoured scenario \citep[e.g.][]{Hilz_et_2013, Dubois_et_2013}.  
Such transitions may or may not involve feedback from the central AGN.  
For instance, based on hydrodynamical cosmological simulations, \citet{Dubois_et_2013} show that the inclusion of jet feedback leads to the quenching of in-situ star formation and switch to the stellar accretion phase. 
As the accreted stars tend to settle in the outer regions, the galaxy's effective radius increases; at the same time, the randomisation of stellar orbits due to satellite accretion also leads to an increased importance of velocity dispersion over regular rotation. 
This induces a transition from compact, rotationally-supported discs into extended, dispersion-dominated ellipsoidal systems.
Both the size growth and the morphological transformation are then attributed to AGN feedback. 
However, the overall evolution is actually driven by the switch in the galaxy growth modes, i.e. from in-situ star formation to satellite accretion, and may be only indirectly caused by AGN feedback.  

In our case, it is the AGN feedback itself that triggers the formation of new stars at outer radii, leading to the development of extended stellar envelopes, which are responsible for the galaxy's size growth. 
The newly born stars initially move on nearly-radial orbits with high radial velocities, in marked contrast to the well-ordered, rotational motion characteristic of disc galaxies. 
Thus structural changes are intrinsically coupled to the size growth, as a result of star formation occurring in the feedback-driven outflow, and the overall evolution may be naturally interpreted in terms of a radially growing stellar distribution. 
Although one usually tends to consider a combination of distinct processes, such as static star formation and external mergers in the `two-phase' scenario, it could be more interesting to consider one single physical scheme in which different aspects of galaxy evolution can be interpreted. 
 
It has long been conjectured that the central black hole regulates the stellar content of its host galaxy, but the exact physical mechanism is still under debate. In our model, the growth of the accreting black hole and the build-up of the host galaxy are inherently coupled through feedback-driven star formation, which provides a direct physical link between the two scales. 
We have seen that the global properties of galaxies, such as characteristic radius and mass, may be directly determined by the effects of radiation pressure on dust. 
In particular, the resulting scaling between radius and mass, of the form $R \propto \sqrt M$, may account for the observed mass-radius relation of early-type galaxies. 
It is interesting to note that the predicted scaling agrees with the size-mass relation of early-type objects, while late-type galaxies are observed to follow a scaling with a different slope. 
This might be a further indication that the important connection is with the spheroidal component. 
In our picture, the chemical evolution is also directly affected by the action of radiation pressure on dust, and the dust content may be an important factor in determining the global shape of galaxies \citep{Santini_et_2014}.
Radiation pressure on dust thus seems to control the large-scale properties of the host galaxy, further setting the black hole-to-bulge mass ratio. 
If this is indeed the case, then the central black hole is actually shaping the basic structure of its host galaxy through AGN feedback. 

Summarising, the connection between AGN feedback and star formation on galactic scales is likely to be more complex than previously thought. 
In particular, the central black hole may not just quench star formation, as generally assumed in the standard negative feedback paradigm. 
In fact, disparate aspects of galaxy evolution may be interpreted within a single framework in which the central black hole plays a major role in shaping its host galaxy through AGN feedback. 

%%%%%%%%%%%%%%%%%%%%%%

\section*{Acknowledgments}

WI acknowledges support from the Swiss National Science Foundation.

\bibliographystyle{mn2e}
\bibliography{biblio.bib}

\begin{thebibliography}{}

\bibitem[\protect\citeauthoryear{{Begelman} \& {Cioffi}}{{Begelman} \&
  {Cioffi}}{1989}]{Begelman_Cioffi_1989}
{Begelman} M.~C.,  {Cioffi} D.~F.,  1989, \apjl, 345, L21

\bibitem[\protect\citeauthoryear{{Bezanson}, {van Dokkum}, {Tal}, {Marchesini},
  {Kriek}, {Franx} \& {Coppi}}{{Bezanson} et~al.}{2009}]{Bezanson_et_2009}
{Bezanson} R.,  {van Dokkum} P.~G.,  {Tal} T.,  {Marchesini} D.,  {Kriek} M.,
  {Franx} M.,    {Coppi} P.,  2009, \apj, 697, 1290

\bibitem[\protect\citeauthoryear{{Buitrago}, {Trujillo}, {Conselice} \&
  {H{\"a}u{\ss}ler}}{{Buitrago} et~al.}{2013}]{Buitrago_et_2013}
{Buitrago} F.,  {Trujillo} I.,  {Conselice} C.~J.,    {H{\"a}u{\ss}ler} B.,
  2013, \mnras, 428, 1460

\bibitem[\protect\citeauthoryear{{Chokshi}}{{Chokshi}}{1997}]{Chokshi_1997}
{Chokshi} A.,  1997, \apj, 491, 78

\bibitem[\protect\citeauthoryear{{Cicone}, {Maiolino}, {Sturm} \& et
  al.}{{Cicone} et~al.}{2014}]{Cicone_et_2014}
{Cicone} C.,  {Maiolino} R.,  {Sturm} E.,    et al. 2014, \aap, 562, A21

\bibitem[\protect\citeauthoryear{{Debuhr}, {Quataert} \& {Ma}}{{Debuhr}
  et~al.}{2011}]{Debuhr_et_2011}
{Debuhr} J.,  {Quataert} E.,    {Ma} C.-P.,  2011, \mnras, 412, 1341

\bibitem[\protect\citeauthoryear{{Dubois}, {Gavazzi}, {Peirani} \&
  {Silk}}{{Dubois} et~al.}{2013}]{Dubois_et_2013}
{Dubois} Y.,  {Gavazzi} R.,  {Peirani} S.,    {Silk} J.,  2013, \mnras, 433,
  3297

\bibitem[\protect\citeauthoryear{{Fabian}}{{Fabian}}{1999}]{Fabian_1999}
{Fabian} A.~C.,  1999, \mnras, 308, L39

\bibitem[\protect\citeauthoryear{{Fabian}}{{Fabian}}{2012}]{Fabian_2012}
{Fabian} A.~C.,  2012, \araa, 50, 455

\bibitem[\protect\citeauthoryear{{Faucher-Gigu{\`e}re} \&
  {Quataert}}{{Faucher-Gigu{\`e}re} \&
  {Quataert}}{2012}]{Faucher-Giguere_Quataert_2012}
{Faucher-Gigu{\`e}re} C.-A.,  {Quataert} E.,  2012, \mnras, 425, 605

\bibitem[\protect\citeauthoryear{{Gaibler}, {Khochfar}, {Krause} \&
  {Silk}}{{Gaibler} et~al.}{2012}]{Gaibler_et_2012}
{Gaibler} V.,  {Khochfar} S.,  {Krause} M.,    {Silk} J.,  2012, \mnras,
  p.~3446

\bibitem[\protect\citeauthoryear{{Genzel}, {Tacconi}, {Gracia-Carpio} \& et
  al.}{{Genzel} et~al.}{2010}]{Genzel_et_2010}
{Genzel} R.,  {Tacconi} L.~J.,  {Gracia-Carpio} J.,    et al. 2010, \mnras,
  407, 2091

\bibitem[\protect\citeauthoryear{{Gomez}, {Krause}, {Barlow}, {Swinyard},
  {Owen}, {Clark}, {Matsuura}, {Gomez}, {Rho}, {Besel}, {Bouwman}, {Gear},
  {Henning}, {Ivison}, {Polehampton} \& {Sibthorpe}}{{Gomez}
  et~al.}{2012}]{Gomez_et_2012}
{Gomez} H.~L.,  {Krause} O.,  {Barlow} M.~J.,  {Swinyard} B.~M.,  {Owen} P.~J.,
   {Clark} C.~J.~R.,  {Matsuura} M.,  {Gomez} E.~L.,  {Rho} J.,  {Besel} M.-A.,
   {Bouwman} J.,  {Gear} W.~K.,  {Henning} T.,  {Ivison} R.~J.,  {Polehampton}
  E.~T.,    {Sibthorpe} B.,  2012, \apj, 760, 96

\bibitem[\protect\citeauthoryear{{Hatch}, {R{\"o}ttgering}, {Miley}, {Rigby},
  {De Breuck}, {Ford}, {Kuiper}, {Kurk}, {Overzier} \& {Pentericci}}{{Hatch}
  et~al.}{2013}]{Hatch_et_2013}
{Hatch} N.~A.,  {R{\"o}ttgering} H.~J.~A.,  {Miley} G.~K.,  {Rigby} E.,  {De
  Breuck} C.,  {Ford} H.,  {Kuiper} E.,  {Kurk} J.~D.,  {Overzier} R.~A.,
  {Pentericci} L.,  2013, \mnras, 436, 2244

\bibitem[\protect\citeauthoryear{{Hilz}, {Naab} \& {Ostriker}}{{Hilz}
  et~al.}{2013}]{Hilz_et_2013}
{Hilz} M.,  {Naab} T.,    {Ostriker} J.~P.,  2013, \mnras, 429, 2924

\bibitem[\protect\citeauthoryear{{Hopkins}, {Hernquist}, {Cox}, {Robertson} \&
  {Krause}}{{Hopkins} et~al.}{2007}]{Hopkins_et_2007}
{Hopkins} P.~F.,  {Hernquist} L.,  {Cox} T.~J.,  {Robertson} B.,    {Krause}
  E.,  2007, \apj, 669, 67

\bibitem[\protect\citeauthoryear{{Ishibashi} \& {Fabian}}{{Ishibashi} \&
  {Fabian}}{2012}]{Paper_1}
{Ishibashi} W.,  {Fabian} A.~C.,  2012, \mnras, 427, 2998

\bibitem[\protect\citeauthoryear{{Ishibashi}, {Fabian} \&
  {Canning}}{{Ishibashi} et~al.}{2013}]{Paper_2}
{Ishibashi} W.,  {Fabian} A.~C.,    {Canning} R.~E.~A.,  2013, \mnras, 431,
  2350

\bibitem[\protect\citeauthoryear{{King}}{{King}}{2003}]{King_2003}
{King} A.,  2003, \apjl, 596, L27

\bibitem[\protect\citeauthoryear{{King}}{{King}}{2010}]{King_2010}
{King} A.~R.,  2010, \mnras, 402, 1516

\bibitem[\protect\citeauthoryear{{Kormendy} \& {Ho}}{{Kormendy} \&
  {Ho}}{2013}]{Kormendy_Ho_2013}
{Kormendy} J.,  {Ho} L.~C.,  2013, \araa, 51, 511

\bibitem[\protect\citeauthoryear{{Krogager}, {Zirm}, {Toft}, {Man} \&
  {Brammer}}{{Krogager} et~al.}{2013}]{Krogager_et_2013}
{Krogager} J.-K.,  {Zirm} A.~W.,  {Toft} S.,  {Man} A.,    {Brammer} G.,  2013,
  ArXiv e-prints

\bibitem[\protect\citeauthoryear{{Krumholz} \& {Thompson}}{{Krumholz} \&
  {Thompson}}{2013}]{Krumholz_Thompson_2013}
{Krumholz} M.~R.,  {Thompson} T.~A.,  2013, \mnras, 434, 2329

\bibitem[\protect\citeauthoryear{{Longhetti}, {Saracco}, {Severgnini}, {Della
  Ceca}, {Mannucci}, {Bender}, {Drory}, {Feulner} \& {Hopp}}{{Longhetti}
  et~al.}{2007}]{Longhetti_et_2007}
{Longhetti} M.,  {Saracco} P.,  {Severgnini} P.,  {Della Ceca} R.,  {Mannucci}
  F.,  {Bender} R.,  {Drory} N.,  {Feulner} G.,    {Hopp} U.,  2007, \mnras,
  374, 614

\bibitem[\protect\citeauthoryear{{Marconi} \& {Hunt}}{{Marconi} \&
  {Hunt}}{2003}]{Marconi_Hunt_2003}
{Marconi} A.,  {Hunt} L.~K.,  2003, \apjl, 589, L21

\bibitem[\protect\citeauthoryear{{McConnell} \& {Ma}}{{McConnell} \&
  {Ma}}{2013}]{McConnell_Ma_2013}
{McConnell} N.~J.,  {Ma} C.-P.,  2013, \apj, 764, 184

\bibitem[\protect\citeauthoryear{{McLure}, {Pearce}, {Dunlop}, {Cirasuolo},
  {Curtis-Lake}, {Bruce}, {Caputi}, {Almaini}, {Bonfield}, {Bradshaw},
  {Buitrago}, {Chuter}, {Foucaud}, {Hartley} \& {Jarvis}}{{McLure}
  et~al.}{2013}]{McLure_et_2013}
{McLure} R.~J.,  {Pearce} H.~J.,  {Dunlop} J.~S.,  {Cirasuolo} M.,
  {Curtis-Lake} E.,  {Bruce} V.~A.,  {Caputi} K.~I.,  {Almaini} O.,  {Bonfield}
  D.~G.,  {Bradshaw} E.~J.,  {Buitrago} F.,  {Chuter} R.,  {Foucaud} S.,
  {Hartley} W.~G.,    {Jarvis} M.~J.,  2013, \mnras, 428, 1088

\bibitem[\protect\citeauthoryear{{Murray}, {Quataert} \& {Thompson}}{{Murray}
  et~al.}{2005}]{Murray_et_2005}
{Murray} N.,  {Quataert} E.,    {Thompson} T.~A.,  2005, \apj, 618, 569

\bibitem[\protect\citeauthoryear{{Newman}, {Ellis}, {Bundy} \& {Treu}}{{Newman}
  et~al.}{2012}]{Newman_et_2012}
{Newman} A.~B.,  {Ellis} R.~S.,  {Bundy} K.,    {Treu} T.,  2012, \apj, 746,
  162

\bibitem[\protect\citeauthoryear{{Patel}, {van Dokkum}, {Franx}, {Quadri},
  {Muzzin}, {Marchesini}, {Williams}, {Holden} \& {Stefanon}}{{Patel}
  et~al.}{2013}]{Patel_et_2013}
{Patel} S.~G.,  {van Dokkum} P.~G.,  {Franx} M.,  {Quadri} R.~F.,  {Muzzin} A.,
   {Marchesini} D.,  {Williams} R.~J.,  {Holden} B.~P.,    {Stefanon} M.,
  2013, \apj, 766, 15

\bibitem[\protect\citeauthoryear{{Prochaska}, {Hennawi} \&
  {Simcoe}}{{Prochaska} et~al.}{2013}]{Prochaska_et_2013}
{Prochaska} J.~X.,  {Hennawi} J.~F.,    {Simcoe} R.~A.,  2013, \apjl, 762, L19

\bibitem[\protect\citeauthoryear{{Rees}}{{Rees}}{1989}]{Rees_1989}
{Rees} M.~J.,  1989, \mnras, 239, 1P

\bibitem[\protect\citeauthoryear{{Roth}, {Kasen}, {Hopkins} \&
  {Quataert}}{{Roth} et~al.}{2012}]{Roth_et_2012}
{Roth} N.,  {Kasen} D.,  {Hopkins} P.~F.,    {Quataert} E.,  2012, \apj, 759,
  36

\bibitem[\protect\citeauthoryear{{Santini}, {Maiolino}, {Magnelli} \& et
  al.}{{Santini} et~al.}{2014}]{Santini_et_2014}
{Santini} P.,  {Maiolino} R.,  {Magnelli} B.,    et al. 2014, \aap, 562, A30

\bibitem[\protect\citeauthoryear{{Shen}, {Mo}, {White}, {Blanton}, {Kauffmann},
  {Voges}, {Brinkmann} \& {Csabai}}{{Shen} et~al.}{2003}]{Shen_et_2003}
{Shen} S.,  {Mo} H.~J.,  {White} S.~D.~M.,  {Blanton} M.~R.,  {Kauffmann} G.,
  {Voges} W.,  {Brinkmann} J.,    {Csabai} I.,  2003, \mnras, 343, 978

\bibitem[\protect\citeauthoryear{{Silk}}{{Silk}}{2005}]{Silk_2005}
{Silk} J.,  2005, \mnras, 364, 1337

\bibitem[\protect\citeauthoryear{{Silk}}{{Silk}}{2013}]{Silk_2013}
{Silk} J.,  2013, \apj, 772, 112

\bibitem[\protect\citeauthoryear{{Silk} \& {Rees}}{{Silk} \&
  {Rees}}{1998}]{Silk_Rees_1998}
{Silk} J.,  {Rees} M.~J.,  1998, \aap, 331, L1

\bibitem[\protect\citeauthoryear{{Sturm}, {Gonz{a}lez-Alfonso}, {Veilleux},
  {Fischer}, {Graci{a}-Carpio}, {Hailey-Dunsheath}, {Contursi}, {Poglitsch},
  {Sternberg}, {Davies}, {Genzel}, {Lutz}, {Tacconi}, {Verma}, {Maiolino} \&
  {de Jong}}{{Sturm} et~al.}{2011}]{Sturm_et_2011}
{Sturm} E.,  {Gonz{a}lez-Alfonso} E.,  {Veilleux} S.,  {Fischer} J.,
  {Graci{a}-Carpio} J.,  {Hailey-Dunsheath} S.,  {Contursi} A.,  {Poglitsch}
  A.,  {Sternberg} A.,  {Davies} R.,  {Genzel} R.,  {Lutz} D.,  {Tacconi} L.,
  {Verma} A.,  {Maiolino} R.,    {de Jong} J.~A.,  2011, \apjl, 733, L16

\bibitem[\protect\citeauthoryear{{Thom}, {Tumlinson}, {Werk}, {Prochaska},
  {Oppenheimer}, {Peeples}, {Tripp}, {Katz}, {O'Meara}, {Brady Ford},
  {Dav{\'e}}, {Sembach} \& {Weinberg}}{{Thom} et~al.}{2012}]{Thom_et_2012}
{Thom} C.,  {Tumlinson} J.,  {Werk} J.~K.,  {Prochaska} J.~X.,  {Oppenheimer}
  B.~D.,  {Peeples} M.~S.,  {Tripp} T.~M.,  {Katz} N.~S.,  {O'Meara} J.~M.,
  {Brady Ford} A.,  {Dav{\'e}} R.,  {Sembach} K.~R.,    {Weinberg} D.~H.,
  2012, \apjl, 758, L41

\bibitem[\protect\citeauthoryear{{van Dokkum}, {Whitaker}, {Brammer}, {Franx},
  {Kriek}, {Labb{\'e}}, {Marchesini}, {Quadri}, {Bezanson}, {Illingworth},
  {Muzzin}, {Rudnick}, {Tal} \& {Wake}}{{van Dokkum}
  et~al.}{2010}]{vanDokkum_et_2010}
{van Dokkum} P.~G.,  {Whitaker} K.~E.,  {Brammer} G.,  {Franx} M.,  {Kriek} M.,
   {Labb{\'e}} I.,  {Marchesini} D.,  {Quadri} R.,  {Bezanson} R.,
  {Illingworth} G.~D.,  {Muzzin} A.,  {Rudnick} G.,  {Tal} T.,    {Wake} D.,
  2010, \apj, 709, 1018

\bibitem[\protect\citeauthoryear{{Zinn}, {Middelberg}, {Norris} \&
  {Dettmar}}{{Zinn} et~al.}{2013}]{Zinn_et_2013}
{Zinn} P.-C.,  {Middelberg} E.,  {Norris} R.~P.,    {Dettmar} R.-J.,  2013,
  \apj, 774, 66

\bibitem[\protect\citeauthoryear{{Zubovas} \& {King}}{{Zubovas} \&
  {King}}{2012a}]{Zubovas_King_2012a}
{Zubovas} K.,  {King} A.,  2012a, \apjl, 745, L34

\bibitem[\protect\citeauthoryear{{Zubovas} \& {King}}{{Zubovas} \&
  {King}}{2012b}]{Zubovas_King_2012b}
{Zubovas} K.,  {King} A.~R.,  2012b, \mnras, 426, 2751

\bibitem[\protect\citeauthoryear{{Zubovas}, {Nayakshin}, {Sazonov} \&
  {Sunyaev}}{{Zubovas} et~al.}{2013}]{Zubovas_et_2013}
{Zubovas} K.,  {Nayakshin} S.,  {Sazonov} S.,    {Sunyaev} R.,  2013, \mnras,
  431, 793

\end{thebibliography}

\label{lastpage}

\end{document}